\pdfoutput=1
\documentclass[aps,prl,twocolumn,superscriptaddress]{revtex4}

\usepackage{graphicx}% Include figure files
\usepackage{longtable}
\usepackage{dcolumn}% Align table columns on decimal point
\usepackage{bm}% bold math
\usepackage{amsmath}
\newcommand{\qp}{$AA'_3B_4$O$_{12}$}
\newcommand{\namno}{(NaMn$_3$)Mn$_4$O$_{12}$}
\newcommand{\camno}{(CaMn$_3$)Mn$_4$O$_{12}$}

\newcommand{\lacamno}{La$_{0.5}$Ca$_{0.5}$MnO$_3$}
\newcommand{\tbmno}{TbMnO$_3$}
\newcommand{\Rcamno}{$R_{0.5}$Ca$_{0.5}$MnO$_3$}

\begin{document}

\title{Commensurate structural modulation in the charge- and orbitally-ordered phase of the quadruple perovskite \namno}

\author{A.~Prodi}
\email[Corresponding Author.~E-mail:]{ <prodi@esrf.fr>}
\thanks{Present address: ESRF, Grenoble, France}
\affiliation{Swiss Light Source, Paul Scherrer Institut, 5232 Villigen, Switzerland}
\author{A.~Daoud-Aladine}
\affiliation{ISIS Facility, STFC Rutherford Appleton Laboratory, Chilton, Didcot, Oxfordshire, OX11 0QX, United Kingdom}
\author{F.~Gozzo}
\thanks{Present address: Excelsus Structural Solutions sprl, 1150 Bruxelles, Belgium}
\author{B. Schmitt}
\affiliation{Swiss Light Source, Paul Scherrer Institut, 5232 Villigen, Switzerland}

\author{O.~Lebedev}
\author{G.~van~Tendeloo}
\affiliation{EMAT, University of Antwerp, 2020 Antwerp, Belgium}

\author{E.~Gilioli}
\author{F.~Bolzoni}
\affiliation{Istituto dei Materiali per Elettronica e Magnetismo, CNR, Area delle Scienze, 43100 Parma, Italy}

\author{H.~Aruga-Katori}
\thanks{Institute of Engineering, Tokyo University of Agriculture and Technology, Koganei, Tokyo 184-8588, Japan}
\author{H.~Takagi}
\affiliation{RIKEN (The Institute of Physical and Chemical Research), 2-1 Hirosawa, Wako, Saitama 351-0198, Japan}

\author{M.~Marezio}
\affiliation{CRETA-CNRS, 25 avenue des Martyrs, 38042 Grenoble, France}

\author{A.~Gauzzi}
\affiliation{IMPMC, UPMC-Sorbonne Universit\'es, CNRS, MNHN, IRD, 4, place Jussieu, 75005 Paris, France}

\date{\today}

\begin{abstract}

By means of synchrotron x-ray and electron diffraction, we studied the structural changes at the charge order transition $T_{CO}$=176 K in the mixed-valence \textit{quadruple} perovskite \namno. Below $T_{CO}$ we find satellite peaks indicating a commensurate structural modulation with the same propagation vector \textbf{q}=(1/2,0,-1/2) of the CE magnetic order that appears at low temperature, similarly to the case of \textit{simple} perovskites like \lacamno. In the present case, the modulated structure together with the observation of a large entropy change at $T_{CO}$ gives evidence of a rare case of full Mn$^{3+}$/Mn$^{4+}$ charge and orbital order consistent with the Goodenough-Kanamori model.
\end{abstract}

% insert suggested PACS numbers in braces on next line

 
% 75.30-m     Intrinsic properties of magnetically ordered materials
% 75.30.Vn    Colossal magnetoresistance
% superstructure

% 61.50.Ks    Crystallographic aspects of phase transformations;
% pressure effects
% 61.66Fn     Structure of specific crystalline solids: Inorganic Compounds

% 61.10.Nz X-ray diffraction
% 61.12.Ld neutron diffraction
% 61.14.-x electron diffraction and scattering

%\keywords{Suggested keywords}
%Use showkeys class option if keyword display desired
% insert suggested keywords - APS authors don't need to do this
%\keywords{}

\maketitle

\textit{Quadruple perovskites} \qp~ recently attracted a great deal of interest for their pronounced charge, spin and orbital orderings \cite{Prodi:04,tak07} and for their promising dielectric \cite{Subramanian:02,Wu:05,Cabassi:06} and multiferroic properties \cite{Zhang:11,joh12,gau13}. These compounds share with \textit{simple perovskites} $AB$O$_3$ a similar pseudocubic network of corner-sharing $B$O$_6$ octahedra but differ for an unusually large tilt of these octahedra stabilized by the Jahn-Teller distortion of the $A'$ site. This significantly reduces the electron hopping and alters the exchange interaction between neighboring $B$ sites. The large tilt reduces to four the number of first nearest oxygen atoms of the $A'$ site, thus preventing oxygen vacancies \cite{gau13}. The disorder inherent in chemically substituted perovskites, such as the prototype system La$_{1-x}$Sr$_x$MnO$_3$ \cite{wol55}, is also absent, for mixed valence of the octahedral $B$ site is achieved by varying separately the valence of the $A$ and $A'$ sites \cite{Marezio:73}.  

\begin{figure}[!b]
\centering
\includegraphics[width=65mm]{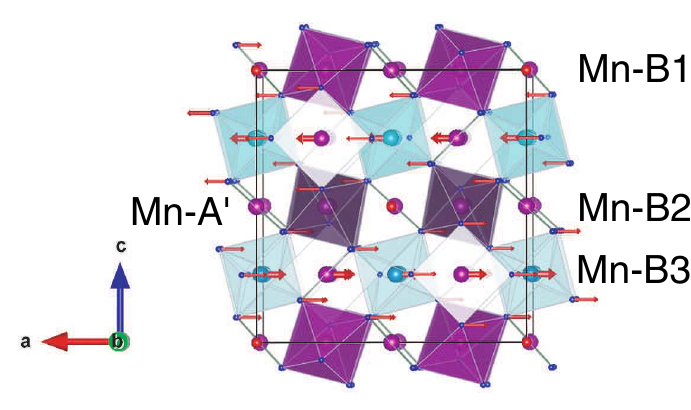}

\caption{\label{fig:struct} (Color online) The monoclinic $C2/m$ structure of \namno~ in the charge-ordered phase below $T_{CO}$. $B$1-3 indicate the non equivalent $B$ octahedral sites of the Mn ions. Mn$^{3+}$O$_6$ and Mn$^{4+}$O$_6$ octahedra are shaded in purple and light blue, respectively. Arrows describe the displacement of the latter octahedra in the modulated structure.}

\end{figure}

In view of these considerations, quadruple perovskites are a model system for studying novel ground states in transition metal oxides in the absence of disorder. Interesting is the case of \namno~ \cite{Marezio:73} owing to the $+3.5$ average valence of the Mn ions in the octahedral $B$ site corresponding to a 0.5 filling of the $e_g$ $d_{x^2-y^2}$-$d_{3z^2-r^2}$ doublet. In this compound, some of us previously reported at $T_{CO}$=176 K an almost full checkerboard-type Mn$^{3+}$/Mn$^{4+}$ charge order (CO) of the $B$ sites in the $ac$-plane, concomitant to a cubic $Im\bar{3}$ to monoclinic $I2/m$ phase transition driven by a pronounced Jahn-Teller distortion of the Mn$^{3+}$O$_6$ octahedra \cite{Prodi:04}. Powder neutron diffraction data indicated a \textit{compression} of the octahedra along the $b$-axis, instead of the zig-zag pattern of \textit{elongated} octahedra in the $ac$-plane, leading to a zig-zag $d_{3z^2-r^2}$ orbital order (OO), characteristic of simple perovskites like \lacamno~ and related compounds at the same filling level. On the other hand, a compression would imply the occupancy of the $d_{x^2-y^2}$ orbital and thus an isotropic exchange interaction between neighboring Mn($B$) ions within the $ac$-plane. According to the Goodenough-Kanamori-Anderson (GKA) rules \cite{Kanamori:59}, this type of interaction would not be compatible with the magnetic CE structure observed at low temperature \cite{Prodi:04}. Shortly before finalizing the present manuscript, a very recent \textit{ab initio} study \cite{str14} has addressed this point by predicting for \namno~ the same pattern of elongated octahedra found previously in \lacamno.  

In order to clarify this point, in this paper we reinvestigated experimentally the low-temperature crystal structure of \namno~ by means of transmission electron and high-resolution powder synchrotron x-ray diffraction. We find a commensurate structural modulation concomitant to the CO transition that confirms the above theoretical prediction and resolves the aforementioned discrepancy between structural distortion and magnetic order by providing a consistent description of the CO-OO order at 0.5 doping in perovskitelike manganites within the GKA model of superexchange.   

\begin{figure}[!t]
\includegraphics[width=65mm]{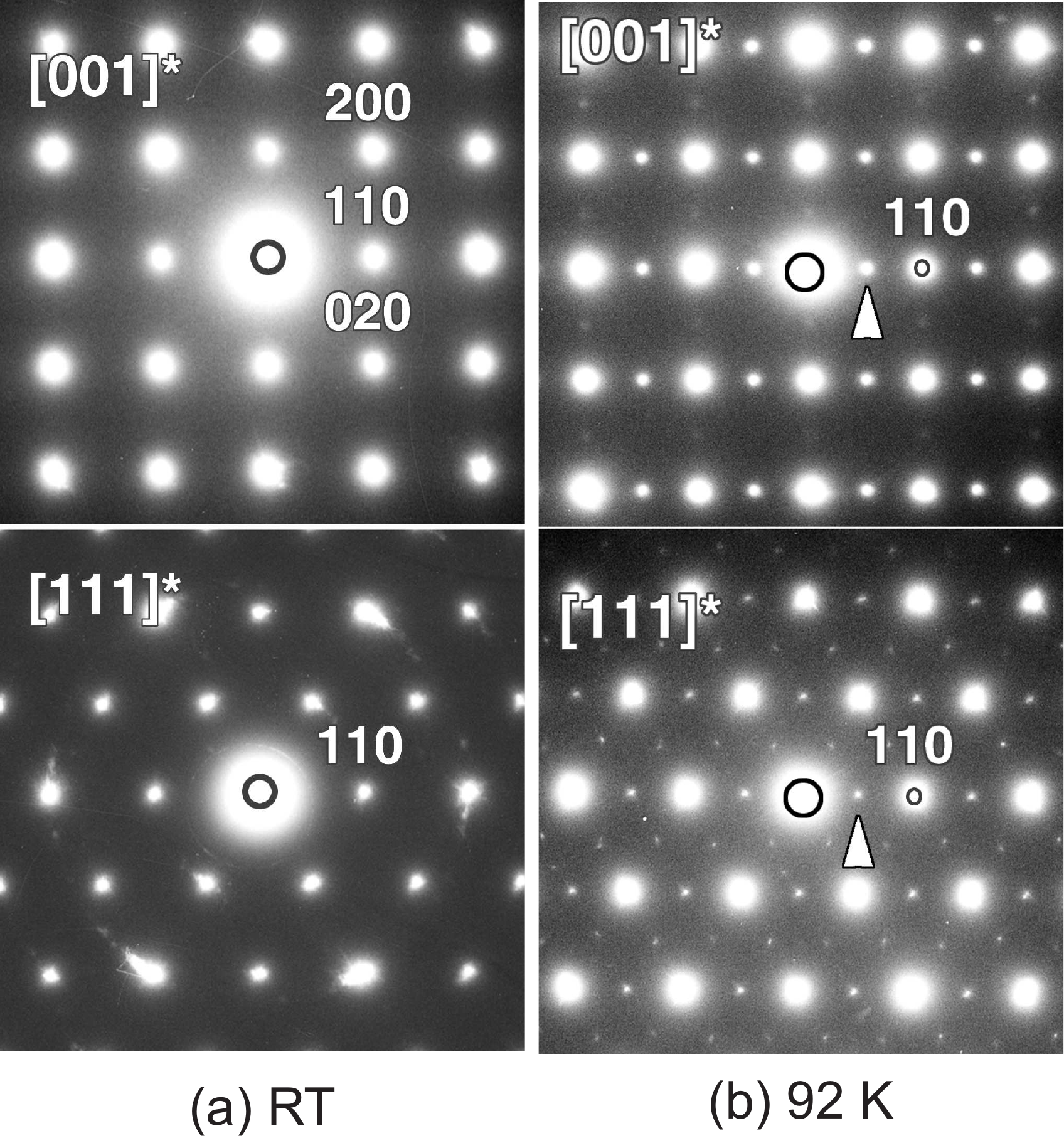}
\caption{\label{fig:edpatt} [001]$^{*}$ and [111]$^{*}$ zone-axis electron diffraction pattern obtained 
(a) above and (b) below the charge-ordering transition in \namno. Arrows indicate the direction of the modulation and indices refer to the cubic $Im\bar{3}$ lattice.}
\end{figure}

For this experiment, we used \namno~ powders synthesized under high pressure, as described elsewhere \cite{Gilioli:05}. X-ray diffraction measurements confirmed that the sample is almost single-phase with $\lesssim$ 4\% weight of Mn$_2$O$_3$ impurities. Electron diffraction experiments were performed at room temperature and at 92 K using a Philips CM20 microscope operating at 200 kV equipped with a low temperature sample stage. High-resolution lattice images detected the presence of grain boundaries but no sign of point defects or microtwins. Thus, the electron diffraction patterns could be recorded on single-domain crystallites. In Fig.~\ref{fig:edpatt}, we compare the patterns along the [001]$^{*}$ and [111]$^{*}$ zone-axis at room temperature and below the structural transition at 92 K for one representative crystallite. The 92 K pattern displays extra spots along the [110] direction, which indicates a commensurate modulation with propagation vector \textbf{q}=(1/2,0,-1/2) corresponding to a doubling of the cell along the $a$- and $c$-axis. The same commensurate \textbf{q} was found in several crystallites.

\begin{figure}
\includegraphics[width=\linewidth]{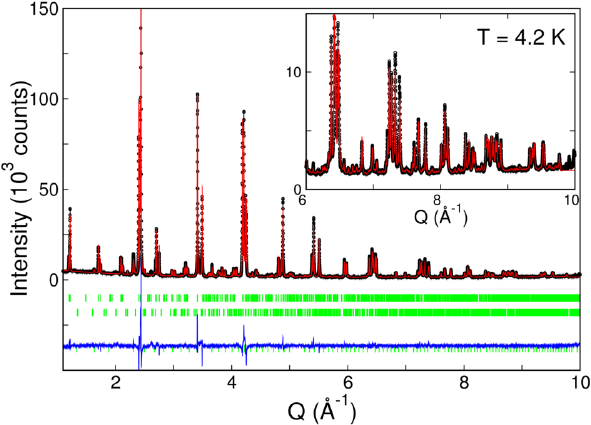}
\caption{\label{fig:sincpat} (Color online) Full-pattern Synchrotron x-ray Rietveld refinement for \namno\ at $T$=4.2 K. The three rows of green ticks mark, from top to bottom, the position of the Bragg peaks of the main $I2/m$ structure, of the superlattice and of the Mn$_2$O$_3$ impurity. The blue line indicates the difference between observed and calculated profile.}
\end{figure}

\begin{figure}
\includegraphics[width=\linewidth]{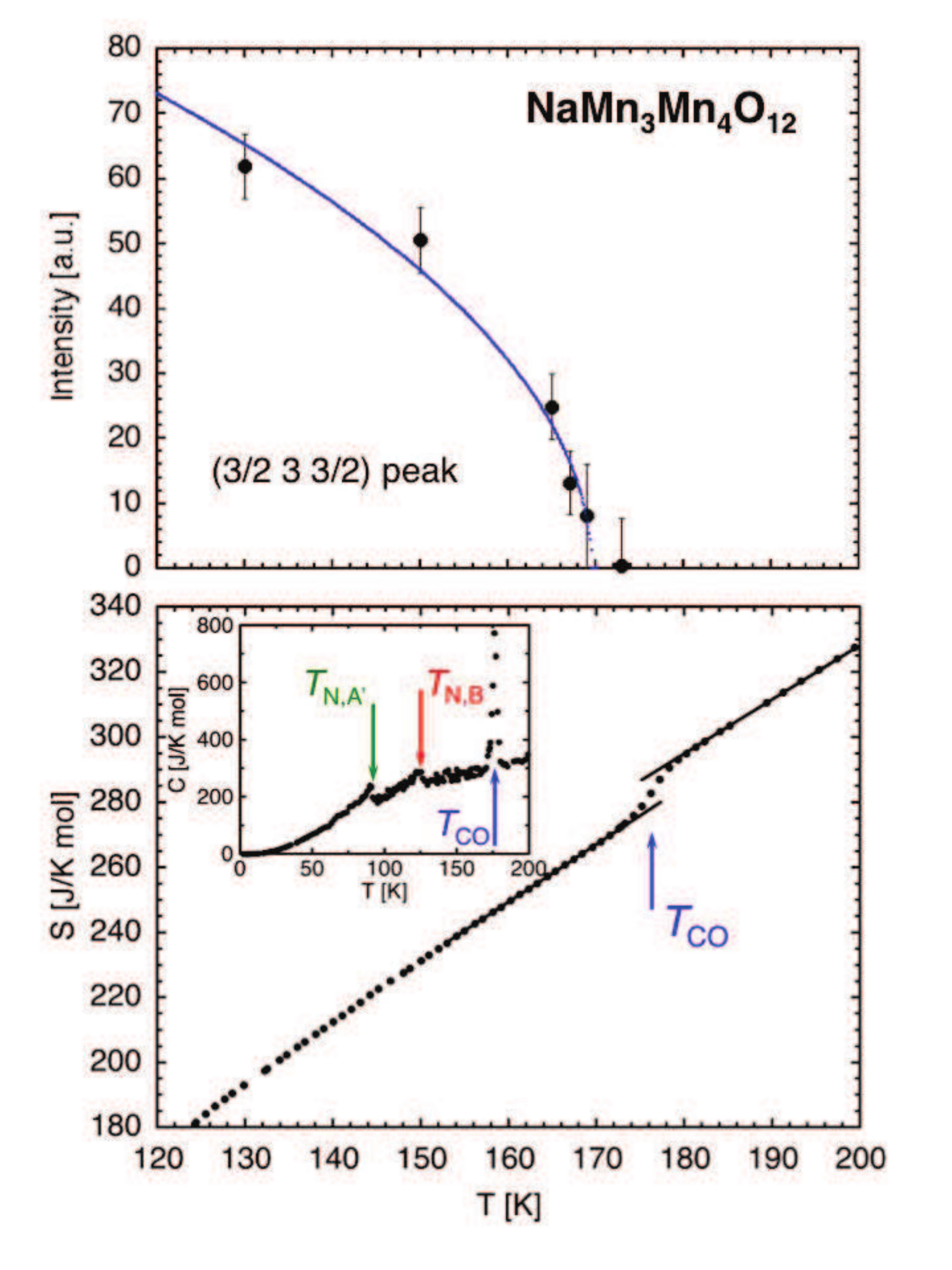}
\caption{\label{fig:sincpat-temp} (Color online) 
Top: temperature dependence of the intensity of the ($\frac{3}{2}$,$3$,$\frac{3}{2}$) superlattice reflection. The solid line is a mean-field fit described by $I(T) \propto (1-T/T_{CO})^{1/2}$. Bottom: dependence of the specific heat, $C(T)$ (inset), and of the entropy, $S(T)$, obtained by direct integration of the $C(T)/T$ data. Note that the divergence in the $C(T)$ curve at $T_{CO}$ appears as a jump in the $S(T)$ curve. $T_{N,A'}$ and $T_{N,B}$ denote the magnetic orderings of the $A'$ and $B$ Mn sublattices.} 
\end{figure}

High-resolution synchrotron x-ray powder diffractograms were collected at several temperatures at the X04SA Beamline \cite{Fabia:04} of the Swiss Light Source in Villigen using a wave length $\lambda$ = 0.708611(1)\AA. A Rietveld refinement confirms the space groups and the lattice parameters reported in the previous neutron diffraction study, specifically the cubic $Im\bar{3}$ to monoclinic $I2/m$ symmetry lowering at $T_{CO}$ \cite{Prodi:04}. In the x-ray diffractogram taken at $T$=4.2 K (see Fig.~\ref{fig:sincpat}), one notes superlattice (SL) reflections indexed with the commensurate propagation vector $\textbf{q}=(1/2,0,-1/2)$, consistent with the electron diffraction data.  We observed a dozen SL reflections that do not overlap with those of the main structure. Their intensities are a few \% of the most intense peaks, while their FWHMs indicate a correlation length of the modulation of $\approx$ 1000 \AA. The temperature evolution of the integrated intensity of the ($\frac{3}{2}$,$3$,$\frac{3}{2}$) reflection shown in Fig.~\ref{fig:sincpat-temp} (a) shows that the modulation occurs at the cubic-monoclinic transition at $T_{CO}$. 

Hereafter we present a model of structural modulation of the $I2/m$ structure previously reported \cite{Prodi:04} which explains the above observations. As in the case of doped manganites, $e.g.$ \lacamno, the modulation has the same periodicity of the CE magnetic structure observed below $T_{N,B}$=125 K, which indicates a simultaneous charge and orbital order. The loss of translational symmetry caused by the modulation implies a lowering of the $I2/m$ symmetry. The group-subgroup path followed by the phase transition corresponds to a symmetry lowering within the double cubic perovskite cell from cubic $Im\bar{3}$ $\rightarrow$ orthorhombic $Immm$ $\rightarrow$ monoclinic $I2/m$, involving $k=$(000) modes and the additional modulation. Such phase transition is required to be of first order in Landau theory \cite{Howard:98}, consistently with the discontinuity of the lattice parameters \cite{Prodi:04} and of the entropy at $T_{CO}$, as seen in Fig.~\ref{fig:sincpat-temp}(a).

The symmetry of the modulated displacement follows the representation $A1+$ of the little group of $I2/m$
with $k = $($\frac{1}{2},0,-\frac{1}{2}$), leading to a supercell with $I2/m$ symmetry ($sg$ 12) and $\mathbf{a}_s=\mathbf{2a}$, $\mathbf{b}_s=\mathbf{b}$ and $\mathbf{c}_s=-\mathbf{a}+\mathbf{c}$. We choose the equivalent $C$-centered monoclinic supercell in the $C2/m$ symmetry with basis vectors $\mathbf{a}_s=\mathbf{a}+\mathbf{c}$, $\mathbf{b}_s=\mathbf{b}$ and $\mathbf{c}_s=-\mathbf{a}+\mathbf{c}$, where \textbf{a}, \textbf{b} and \textbf{c} are the basis vectors of the non-modulated $I2/m$ monoclinic structure and the subscript $s$ stays for supercell.

\begin{figure}[!ht]
 \centering
\includegraphics[width=80mm]{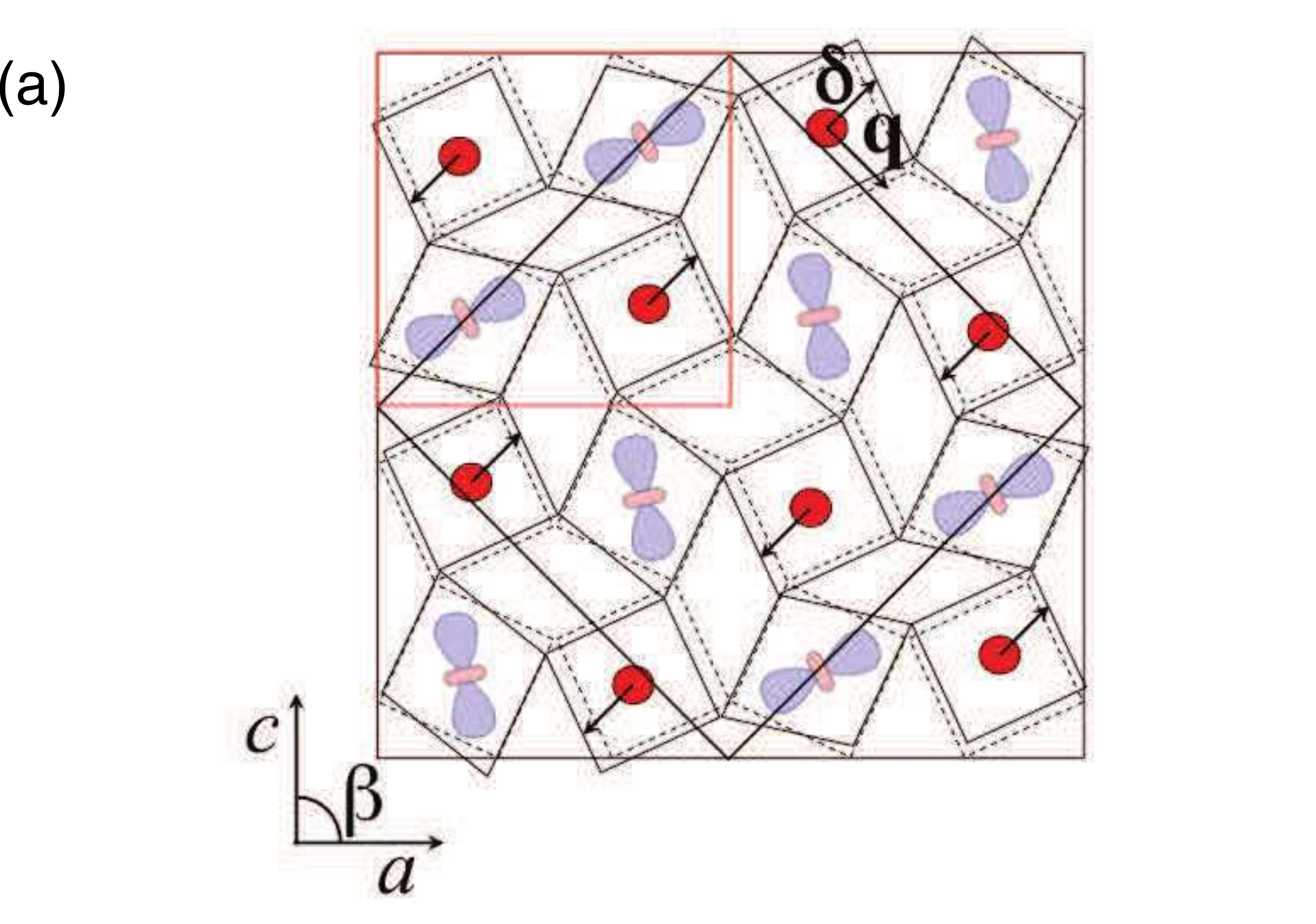}
%\vspace{0.1cm}
%\includegraphics[width=55mm]{./fig5b}

\vspace{0.3cm}

\includegraphics[width=85mm]{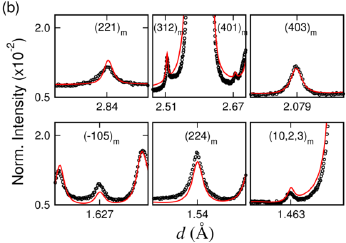}
\caption{\label{fig:modulaz} (Color online) (a) Model of structural modulation in the charge-ordered phase. The small red square on the top left corner indicates the undistorted cubic $Im\bar{3}$ cell. The large black square rotated by 45$^{\circ}$ indicates the supercell. The undistorted and distorted octahedra are sketched by using broken and continuous black lines, respectively. Arrows indicate the shear displacement by $\delta$ of the rows of Mn$^{4+}$O$_6$ octahedra along the direction perpendicular to the propagation vector, \textbf{q}. The $e_g$ $3z^2-r^2$ orbitals in the elongated Mn$^{3+}$O$_6$ octahedra are schematically shown. (b) Agreement of the refined model with the strongest non-overlapping superlattice reflections. Intensities are normalized to the most intense Bragg peak; indices refer to the $C2/m$ monoclinic cell.}
\end{figure}

In principle, the modulated structure could be refined by using the above supercell in the $C2/m$ symmetry; in practice, this is not feasible because of the large (27) number of atomic parameters and the difficulty of modelling the asymmetric peak profile. Therefore, we adopted a phenomenological model similar to that proposed by Radaelli \textit{et al.} \cite{Radaelli:97} for the counterpart compound with simple perovskite structure \lacamno. The distortion pattern, shown in Fig.~\ref{fig:modulaz}(a), consists of a transverse displacement, $\delta$, of the Mn$^{4+}$O$_6$ octahedra in the $ac$-plane along the direction perpendicular to \textbf{q}. The displacements are assumed to be rigid because the above octahedra are not Jahn-Teller active. This leads to a shear displacement of the rows of these octahedra along opposite directions perpendicular to \textbf{q}. The displacement leads to the aforementioned zig-zag pattern of elongated Mn$^{3+}$O$_6$ octahedra. In order to take into account the displacement of the Mn $A'$ ions, it is noted that, while the Na($A$)-O bond is ionic and isotropic due to the icosahedral coordination, the Mn($A'$)-O bond has semi-covalent character and the Mn($A'$) ion connecting four octahedral corners forms a square-shaped plaquette. We assume the displacement of the plaquettes to be rigid because a non-planar coordination of the Mn$^{3+}$ ion is not favoured energetically due to the occupation of the $t_{2g}$ orbitals. Thus, the coordinated displacement of the plaquettes and of the octahedra takes place in an accordion-plaited fashion.

The above model was successfully refined by employing the modulated structure description implemented within the FullProf package \cite{Fullprof:90}. The refinement converged to a value $\delta$=0.06730(55)\AA\ with a reliability factor, $R \approx 6.3$. In Fig.~\ref{fig:modulaz}(b), note the good agreement for the superlattice reflections. By using the FINDSYM program \footnote{FINDSYM: H. T. Stokes and D. M. Hatch, J. of Appl. Cryst. 38, 237 (2005).}, the symmetry of the refined modulated structure was determined to be $C2/m$, with $a$=10.35735(15), $b$=7.19562(11), $c$=10.45472(5) and $\beta$=90.0206(11)$^{\circ}$. In the modulated structure, the octahedral Mn($B$) site splits into one Mn$^{4+}$ site with $8j$ general position and two Mn$^{3+}$ sites ($4e-4f$). The analysis of the refined Mn-O bond lengths (see Table~\ref{tab:table1}) shows that the staggered distribution of the long Mn-O bonds in the distorted octahedra Mn($B$1) and Mn($B$2) are consistent not only with the CE-type of magnetic order, but also with the picture of static charge disproportionation stabilized by the superexchange interaction, in agreement with the non modulated structure proposed earlier. Indeed, a bond-valence-sum analysis confirms that the valence of the undistorted Mn site is 4+, while that of the distorted Mn sites is 3.18 and 3.34 respectively, similar to the values obtained previously for the average structure. The MnO$_4$ plaquette sites remain in square-planar coordination, with $d$(Mn-O)= 1.865 \AA\, 1.919 \AA\ and 1.953 \AA, respectively. Note that the present structural model basically coincides with the aforementioned \textit{ab initio} calculations \cite{str14}. Alternative models, such as the Zener-polaron model proposed earlier for manganites at 0.5 doping \cite{Aziz:02}, which considers the ordering of Mn dimers formed by the double exchange interaction, are not consistent with the present results.

\begin{table}[!ht]
\caption{\label{tab:table1}Mn-O bond distances for the three Mn $B1-3$ octahedral sites shown in Fig. 1 resulting from the $C2/m$ refined modulated structure at $T$=4.2 K. We also report the average distances, the bond valence sum (BVS) and the distortion parameter, $\Delta=\frac{1}{6}\Sigma_{i=1}^6 \vert\frac{d_{\text{MnO}_i}-\langle d_{\text{MnO}} \rangle}{\langle d_{\text{MnO}} \rangle}\vert ^2$, for these sites.}
\vspace{0.2cm}
\begin{ruledtabular}
\begin{tabular}{cccc}
Mn $B$ site   & Mn($B$1) & Mn($B$2) & Mn($B$3)  \\
\hline \\ [-1.5ex]
$d$(Mn-O)/\AA & 1.922  & 1.922  & 1.929  \\
              & 1.922  & 1.922  & 1.929  \\
              & 2.114  & 1.955  & 1.883  \\
              & 2.114  & 1.955  & 1.883  \\
              & 1.974  & 2.067  & 1.900  \\
              & 1.974  & 2.067  & 1.900   \\
$\langle d$(Mn-O)$\rangle$/\AA & 2.003 & 1.981  & 1.904  \\
BVS           & 3.18   & 3.34  & 3.99 \\
$\Delta$($\times$10$^{-4}$)& 16.33 & 9.84 & 1.00 \\
\end{tabular}
\end{ruledtabular}
\end{table}

The picture of static charge disproportionation is further supported by a specific heat study as a function of temperature. For these measurements, we used a MagLab microcalorimeter (Oxford Instruments) that employs a thermal relaxation method. In order to exclude any effects arising from granularity or secondary phases, the measurements have been carried out on a bunch of small single crystals. Fig.~\ref{fig:sincpat-temp} (b) show the specific heat data in the vicinity of the CO transition and the corresponding entropy difference, $\Delta S_{CO}$. Owing to the negligible latent heat at the transition, $\Delta S_{CO}$ was obtained by a straightforward integration of the specific heat data, as described in ref. \cite{PhysRevB.71.024426} for the doped perovskite Pr$_{0.4}$Ca$_{0.6}$MnO$_3$. From Fig.~\ref{fig:sincpat-temp}, in the present case, we find $\Delta S_{CO}=11 \pm 1$ mJ K$^{-1}$ mol$^{-1}$, comparable to the value 4$R\rm{ln}2$=23 mJ K$^{-1}$ mol$^{-1}$ expected for an ideal order-disorder phase transition of an equal proportion of Mn$^{3+}$ and Mn$^{4+}$ ions within a square lattice containing 4 sites, which corresponds to the \qp~ formula unit. In Pr$_{0.4}$Ca$_{0.6}$MnO$_3$, a value 40 \% smaller was obtained \cite{PhysRevB.71.024426}, which indicates that the charge ordering is less pronounced in this doped manganite.

The present results may be compared with those reported on \camno~ \cite{Bochu:74,Bochu:80,Prezioso:04,Perks:2012} that also displays at $T_{OO}$=250 K a modulation of a rhombohedrally distorted structure concomitant to a 3:1 ordering of the Mn$^{3+}$/Mn$^{4+}$ $B$ ions. The non modulated structure was also reported to contain an apically compressed Mn$^{3+}$ site \cite{Bochu:80}. However, contrary to the present case, in \camno, $T_{OO}$ is much lower than $T_{JT}$=440 K at which the Jahn-Teller-driven distortion occurs. Moreover, the modulation is incommensurate and of spiral type. Further theoretical studies may account for the unique type of commensurate and full charge ordering in \namno. We argue that the existence of two distinct superexchange interaction paths, $A'$-O-$B$ and $B$-O-$B$ with unusually small angles, $\psi \approx 110^{\circ}$ and $\approx 135^{\circ}$, respectively, is expected to enhance the next-nearest-neighbor interactions, as in \tbmno \cite{PhysRevLett.97.227204}.

In conclusion, by means of a high-resolution structural study, we have given experimental evidence for a commensurate structural modulation that consistently accounts for the simultaneous CO/OO ordering in \namno\ within the framework of the GKA scheme. Unique properties of the above CO/OO ordered phase are (i) the commensurability of the structural modulation described by the same propagation vector of the CE magnetic order; (ii) the almost full static Mn$^{3+}$/Mn$^{4+}$ charge disproportionation; (iii) the coincidence of the CO/OO ordering with the Jahn-Teller distortion of the Mn$^{3+}$ site. These properties are ascribed to the absence of disorder caused by chemical substitutions or oxygen defects, which rules out the role of the double exchange interaction, contrary to the case of \Rcamno \cite{VanDenBrink:99}. This simple phenomenology should foster theoretical studies that may convincingly account for the stability of the above CO/OO orderings and for the functional properties hitherto reported. For instance, it would be interesting to unveil the link between the unusually large tilt of the octahedral network with the enhanced improper ferroelectricity found in the related compound \camno~ \cite{Zhang:11,joh12}.   

We thank G. Calestani, C. Mazzoli, B. Patterson and P.G. Radaelli for useful discussions. A.P. acknowledges financial support from the Fondazione A. Della Riccia and from the Paul Scherrer Institut.

\bibliography{manga-new}

\end{document}